\documentclass[twocolumn,pra,tightenlines,showpacs,floatfix]{revtex4}
\usepackage{amsmath}
\usepackage{graphicx}
\usepackage{amssymb}
\usepackage{esint}

\renewcommand{\vec}[1]{\boldsymbol #1}

\DeclareMathOperator{\Trace}{Tr}\DeclareMathOperator{\STr}{STr}
\DeclareMathOperator{\erf}{erf}

\begin{document}

\title{Convergence of a renormalization group approach to dimer-dimer scattering}

\author{Michael C. Birse$^{1}$, Boris Krippa$^{1,2}$, Niels R. Walet$^{1}$}

\affiliation{$^{1}$School of Physics and Astronomy, The University of Manchester,
Manchester, M13 9PL, UK\\
 $^{2}$Institute for Theoretical and Experimental Physics, Moscow,
117259, Russia}

\date{\today}
\begin{abstract}
We study the convergence of a functional renormalisation group technique
by looking at the ratio between the fermion-fermion scattering length
and the dimer-dimer scattering length for a system of nonrelativistic
fermions. We find that in a systematic expansion in powers of the
fields there is a rapid convergence of the result that agrees with
know exact results.
\end{abstract}


\maketitle

\section{introduction}

The atomic physics of ultra-cold Fermi gases is one of the places
where we can make a detailed link few-particle and many-body physics.
In fermionic systems we can trace the effect of an attractive force
from the bound states or resonances in few-body systems to the pairing
occurring in the many-body sector. At low energy and in cold gases
this physics is governed by a single parameter, the central $S$-wave
scattering length $a_{F}$, determining the scattering at threshold.
We can also tune the value of this scattering length making use of
recent advances using Feshbach resonances. For negative scattering
length the gas is in the weak-coupling BCS state. For positive values
of $a_{F}$ bound states of two fermions--{}``dimers''--form and
these can lead to a Bose-Einstein condensate (BEC) \cite{Zwe}. The
size of dimers is determined by the fermion-fermion scattering length
and their binding energy is of order $1/a_{F}^{2}$.

If we concentrate on the case of deeply-bound dimers, then for a sufficiently
dilute and cold gas of dimers the main dynamical quantity characterising
their interaction is now the dimer-dimer scattering length $a_{B}$,
which is a induced by the fermionic scattering. The exact relation
between dimer-dimer and fermion-fermion scattering lengths $a_{B}=0.6a_{F}$
was established in Ref.~\cite{Shl} by solving the Schrödinger equation
for two composite bosons interacting with an attractive zero-range
potential. Unfortunately, this method is difficult to extend to the
many-body case. Therefore, it is useful to study the ratio $a_{B}/a_{F}$
in an approach which can be used both for few and many-body problems.

One such method is the functional or {}``exact'' renormalisation
group (ERG) for the one-particle irreducible effective action, which
is the approach applied in this paper to calculate $a_{B}$. (For
reviews, see Refs.~\cite{BTW,DMT}). This technique has been previously
used to study a variety of physical systems, from systems of nonrelativistic
fermions \cite{Kri,Kri2,Diehl,Bir,DKS,Flo,Flo2} and bosons \cite{SandM}
to quark models \cite{JW} and gauge theories \cite{LPG}. It is based
on the scale-dependent quantum effective action $\Gamma_{k}$, where
the action at scale $k$ contains the effects of field fluctuations
with momenta $q$ larger than $k$ only. In the limit $k\rightarrow0^{+}$
all fluctuations are included and the full effective action is recovered.
In practice the scale dependence is introduced by a set of $k$-dependent
cutoff functions $R(q)$, which suppress the effect of modes with
$q<k$ in the path integral for the action by giving them a large
$k$ dependent mass. The minimal conditions satisfied by the functions
$R(q)$ are then that they should vanish in the limit $k\rightarrow0^{+}$
and scale like $k^{\alpha}$ with $\alpha>0$ when $k\rightarrow\infty$
. 

With this prescription the average effective action at very large
$k$ converges to the classical action of the theory---here nonrelativistic
fermions with a local zero-range interaction. Since one can show that
we get the effective action for the theory as $k\rightarrow0^{+}$,
the endpoint at $k=0$ of the exact solution of the functional RG
equation should be independent of the choice of cutoff. However, in
practice, truncations of the action inevitably lead to some cutoff
dependence of the results. We can use this dependence as one possible
measure of the quality of the truncation. With this tool, we shall
analyse the convergence of a low-energy long-wavelength expansion
as we increase the complexity of the many-body truncation in the effective
action.

As discussed in great detail in the literature, e.g. Ref.~\cite[Chapter 16]{Wein2},
the idea of the (quantum) effective action $\Gamma$ is to introduce
an object that generalises the concept of the classical action of
a field theory to include all quantum effects, but still depends on
a set of classical external fields (the dual of the usual external
sources in the partition function). The ground state is the minimum
of the action, and we can generate all Green functions by appropriate
derivatives of $\Gamma$. Apart from for very simple models, it is
very hard to explicitly evaluate the full effective action, which
is given by a complicated path integral. The ERG technique used here
gives a scale-dependent mass-gap to the low-momentum modes in the
path integral, so that their fluctuations are suppressed. The action
then flows from the trivial case of all fluctuations suppressed (scale
$k$ is infinite) to the full quantum effective action for zero scale.
Since this approach in principle incorporates the full quantum dynamics
of the underlying theory, it is called {}``exact''.

The flow of the effective action satisfies the functional differential
equation \cite{BTW} \begin{equation}
\partial_{k}\Gamma=-\frac{i}{2}\,\STr\left[(\partial_{k}R)\,(\Gamma^{(2)}-R)^{-1}\right].\label{eq:ERG1}\end{equation}
 where $\Gamma^{(2)}$ is the second functional derivative with respect
to the fields, and the cutoff functions in the mass-like term $R(k)$
drive the RG evolution. The operation $\STr$ denotes the supertrace
\cite{Ma} taken over energy-momentum variables and internal indices
and is defined by \begin{equation}
\STr\left(\begin{array}{cc}
A_{BB} & A_{BF}\\
A_{FB} & A_{FF}\end{array}\right)=\Trace(A_{BB})-\Trace(A_{FF}).\end{equation}

The evolution equation for the average effective action thus has a
one-loop structure, but contains a fully dressed, scale-dependent
propagator $(\Gamma^{(2)}-R)^{-1}$. Thus, despite its apparently
simple form, Eq.~(\ref{eq:ERG1}) is actually a complex functional
differential equation. In the absence of general methods to numerically
solve such equations we must resort to approximations. One common
approach is to parametrise the effective action with a small number
of terms, which turns the evolution equation into a system of coupled
ordinary differential equations for the numerical coefficient of each
term. These equations can then be solved numerically. Here we study
possible truncations for fermionic few-body systems, and our choice
of ansatz for the action is motivated by both ERG studies of many-body
systems \cite{Kri,Diehl} and effective field theories (EFTs) for
few-fermion systems \cite{EFT}. The technique we use is similar to
the one used in Ref.~\cite{Bir} and Ref.~\cite{Kri2}. The parametrisation
of the effective action is given by a local expansion in fundamental
and induced fields, with a simple gradient expansion for non-local
terms.

There are a number of questions about the approximations made in writing
down the action. We shall try and answer two of these in this paper:
Firstly, does the low-energy truncation, where we expand to first
order in energy and to first order in field-gradient squared converge?
Secondly, are these results independent of the choice of cut-off function?
These two questions are not independent: If we have a converged truncation
of the effective action, we \emph{must} have independence of the cut-off
function. Reversing this, we can hope that the dependence on the cut-off
can be used as one of the signals of convergence.

The renormalisation functions $R(q)$ is not completely arbitrary:
as the cutoff scale tends to infinity, we demand that the action be
purely fermionic, containing a contact two-body interaction without
derivatives. Such interactions are of interest also for interacting
nucleons \cite{EFT}. In both these cases, it is convenient to add
an auxiliary composite boson field (which is really the dimer field)
to the problem by making a Hubbard-Stratonovich transformation. This
then replaces the two-body interaction by a Yukawa-type coupling between
the fermions and the auxiliary boson. At the extreme end of the scale,
the boson is not dynamic (there is no energy or momentum dependence
of this field in the action), but a kinetic term for the boson is
generated in the RG evolution.

\section{Building blocks}

We define the low-energy effective action following previous work
(see \cite{Kri,Diehl,Bir,Kri2,SandM}) as an expansion in the fermion
fields $\psi$--a two-component spinor field, where the index denotes
fermion spin-- and the scalar dimer boson field $\phi$ obtained from
bosonising the bilinear fermion field $\psi^{{\rm T}}(x)\sigma_{2}\psi(x)$
\begin{align}
 & \Gamma[\psi,\psi^{\dagger},\phi,\phi^{\dagger},k]\nonumber \\
 & =\int d^{4}x\,\Biggl[\int d^{4}x'\,\phi^{\dagger}(x)\Pi(x,x';k)\phi(x')\nonumber \\
 & \quad+\psi^{\dagger}(x)\left(i\partial_{t}+\frac{1}{2M}\,\nabla^{2}\right)\psi(x)\nonumber \\
 & \quad-\frac{i}{2}\, g\biggl(\psi^{{\rm T}}(x)\sigma_{2}\psi(x)\phi^{\dagger}(x)-\psi^{\dagger}(x)\sigma_{2}\psi^{\dagger{\rm T}}(x)\phi(x)\biggr)\nonumber \\
 & \quad-\frac{1}{2}\, u_{2}\,\Bigl(\phi^{\dagger}(x)\phi(x)\Bigr)^{2}\nonumber \\
 & \quad-\lambda\,\phi^{\dagger}(x)\phi(x)\psi^{\dagger}(x)\psi(x)\nonumber \\
 & \quad-\frac{i}{4}\, g'\Bigl(\phi^{\dagger}(x)\phi(x)\Bigr)\nonumber \\
 & \qquad\biggl(\psi^{{\rm T}}(x)\sigma_{2}\psi(x)\phi^{\dagger}(x)-\psi^{\dagger}(x)\sigma_{2}\psi^{\dagger{\rm T}}(x)\phi(x)\biggr)\nonumber \\
 & \quad-\frac{1}{4}\,\nu\,\Bigl(\phi^{\dagger}(x)\phi(x)\Bigr)\psi^{\dagger}(x)\sigma_{2}\psi^{\dagger{\rm T}}(x)\psi^{{\rm T}}(x)\sigma_{2}\psi(x)\Biggr].\label{eq:ansatz-1}\end{align}
In contrast with the many-body case, in the few-body sector there
is no wavefunction and mass renormalisation of the fermion field.
The quantity $\Pi(x,x',k)$ is the scale-dependent boson self-energy
and $u_{2}$ parametrises the boson-boson interaction which is generated
by the evolution. The latter is equivalent to a four-body interaction
in terms of the underlying fermions. The term proportional to $\lambda$
describes the fermion-dimer scattering (three body in the fermions).
The final two terms are four-body again: the term proportional to
$g'$ describes the scattering of two dimers, where one dimer breaks
up into two fermions, and finally the $\nu$ term describes the scattering
of two fermions from a dimer, without changing character. We have
not included the local two-fermion interaction, which has zero coupling
constant due to our choice of starting model at large (infinite scale),
where only $g$ is nonzero. 

The evolution of the boson self-energy is given by \begin{equation}
\partial_{k}\Pi(x,x',k)=\frac{\delta^{2}}{\delta\phi(x')\delta\phi^{\dagger}(x)}\,\partial_{k}\Gamma|_{\phi=0},\end{equation}
 although from now on we shall express all evolution in momentum space.
Note that only fermion loops contribute to the evolution of the boson
self-energy in vacuum. 

Using a gradient expansion of the action, we can define boson wave-function
and mass renormalisation factors by \begin{equation}
Z_{\phi}(k)=\frac{\partial}{\partial P_{0}}\,\Pi(P_{0},\vec{P},k)\Biggr|_{P_{0}=\mathcal{E}_{D},\vec{P}=0},\end{equation}
 and \begin{equation}
\frac{1}{4M}\, Z_{m}(k)=-\frac{\partial}{\partial P^{2}}\,\Pi(P_{0},\vec{P},k)\Biggr|_{P_{0}=\mathcal{E}_{D},\vec{P}=0},\end{equation}
 where $\mathcal{E}_{D}=-1/(Ma_{F}^{2})$ denotes the bound-state
energy of a pair of fermions. It is relatively straightforward to
show that $Z_{m}(0)=Z_{\phi}(0)$, independent of the cut-off function.
There are very few cut-off functions that preserve Galilean invariance
to quadratic order in momenta, but one favourite sharp cut-off \cite{Litim}
\begin{equation}
R_{F}({\vec{q}},k)=\frac{k^{2}-q^{2}}{2M}\,\theta(k-q),\label{eq:LitimCutoff}\end{equation}
can be shown to preserve the identity $Z_{\phi}(k)=Z_{m}(k)$. Such
sharp cut-off functions are difficult to apply in medium, since the
evolution of $Z_{m}$ will contain ambiguous terms containing $\delta$
functions and their derivatives arising from the first and second
derivative of the cut-off function. Such difficulties can be bypassed
here since $\Pi$ can be evaluated directly. We first impose the boundary
condition that the scattering amplitude in the physical limit $k\rightarrow0$
reproduces the fermion-fermion scattering length, \begin{equation}
\frac{1}{T(p)}=\frac{1}{g^{2}}\,\Pi(P_{0},P,0)=\frac{M}{4\pi a_{F}}\,.\end{equation}
 Here $P_{0}\ (P)$ denotes the total energy (momentum) flowing through
the system and $p=\sqrt{2MP_{0}-P^{2}/2}$ is the relative momentum
of the two fermions. Integrating the resulting ERG equation gives
\cite{Bir} \begin{eqnarray}
\Pi(P_{0},P,k) & = & \frac{g^{2}M}{4\pi^{2}}\biggl[-\frac{4}{3}\, k+\frac{\pi}{a_{F}}+\frac{16}{3k}\left(MP_{0}-\frac{P^{2}}{2}\right)\nonumber \\
 &  & \qquad\qquad-\frac{P^{3}}{24k^{2}}+...\biggr].\end{eqnarray}
This shows that Galilean invariance is preserved only to lowest order.

The complexity of sharp cut-off functions in medium implies that it
is of interest to study smooth cut-offs, where both $Z$'s must now
be calculated independently. A suitable set of smooth functions can
be parametrised as %
\footnote{Note that we use $k^{2}$ as a prefactor here; using $k^{2}-q^{2}$
instead leads to singularities in some integrals.%
}\begin{align}
R_{F}({\vec{q}},k) & =\frac{k^{2}}{2M}\,\theta_{\sigma}(q,k),\nonumber \\
\theta_{\sigma}(q,k) & =\frac{\erf\left((-q/k+1)/\sigma\right)+\erf\left((q/k+1)/\sigma\right)}{2\erf(1/\sigma)}.\label{eq:SmoothCutoff}\end{align}

\subsection{Mean-field}

In vacuum, it is easy to show that\begin{align*}
Z_{\phi}(k) & =\frac{g^{2}}{4}\int\frac{d^{3}{\vec{q}}}{(2\pi)^{3}}\,\frac{1}{\left(E_{FR}(\vec{q},k)-\mathcal{E}_{D}/2\right)^{2}},\\
Z_{m}(k) & =\frac{g^{2}}{6}\int\frac{d^{3}{\vec{q}}}{(2\pi)^{3}}\,\frac{1}{6q\left(E_{FR}(\vec{q},k)-\mathcal{E}_{D}/2\right)^{2}}\\
 & \qquad\times\left(2\partial_{q}E_{FR}(\vec{q},k)+q\partial_{qq}E_{FR}(\vec{q},k)\right).\end{align*}
Thus\[
Z_{\phi}(0)=Z_{m}(0)=\frac{g^{2}a_{F}}{8\pi}.\]

The evolution of the boson-boson scattering amplitude follows from
\begin{equation}
-\frac{2}{(2\pi)^{4}}\partial_{k}u_{2}(\mathcal{E}_{D},k)=\frac{\delta^{4}}{\delta\phi^{2}(\mathcal{E}_{D},0)\delta\phi^{\dagger2}(\mathcal{E}_{D},0)}\,\partial_{k}\Gamma|_{\phi=0}.\end{equation}
 The driving term of this equation can be separated into fermionic
and bosonic contributions containing $\partial_{k}R_{F}$ and $\partial_{k}R_{B}$,
respectively. We first look at the {}``mean-field'' result, where
bosonic contributions are neglected. The evolution of $u_{2}$ is
then given by \begin{equation}
\partial_{k}u_{2}=-\frac{3g^{4}}{4}\int\frac{d^{3}{\vec{q}}}{(2\pi)^{3}}\,\frac{\partial_{k}R_{F}}{\left[(E_{FR}(\vec{q},k)-\mathcal{E}_{D}/2\right)]^{4}},\label{eq:u2F}\end{equation}
 where \begin{equation}
E_{FR}(\vec{q},k)=\frac{1}{2M}\, q^{2}+R_{F}(q,k).\end{equation}
Equation (\ref{eq:u2F}) is integrable, and we find with $u_{2}(\infty)=0$
that \begin{align}
u_{2}(k) & =\frac{3g^{4}}{16}\int\frac{d^{3}{\vec{q}}}{(2\pi)^{3}}\,\frac{1}{\left[(E_{FR}(\vec{q},k)-\mathcal{E}_{D}/2\right)]^{3}},\\
u_{2}(0) & =\frac{1}{16\pi}\, M^{3}g^{4}a_{F}^{3}.\end{align}
 The scattering amplitude at threshold is \begin{equation}
T_{BB}=\frac{8\pi}{2M}\, a_{B}=\frac{2u_{2}(0)}{Z_{\phi}^{2}}=\frac{8\pi a_{F}}{M},\end{equation}
 giving the well-known mean-field result $a_{B}=2a_{F}$ \cite{Ha}
which is far from the exact value $a_{B}=0.6a_{F}$ \cite{Shl}. This
implies that beyond-mean-field effects such as dimer-dimer rescattering
must be considered.

\subsection{Boson and mixed loops}

To include such effects we must take into account the loops containing
boson propagators. After some algebra, we find the driving term \begin{equation}
\partial_{k}u_{2}|_{B}=\frac{u_{2}^{2}(k)}{2Z_{\phi}(k)}\int\frac{d^{3}{\vec{q}}}{(2\pi)^{3}}\,\frac{\partial_{k}R_{B}}{\left[E_{BR}(\vec{q},k)-\mathcal{E}_{D}\right]^{2}},\end{equation}
 where \begin{equation}
E_{BR}(\vec{q},k)=\frac{Z_{m}(k)}{4M}\, q^{2}+u_{1}(k)+R_{B}(q,k)\end{equation}
 and \begin{equation}
u_{1}(k)=-\Pi(\mathcal{E}_{D},0,k).\end{equation}
In the case of a sharp cutoff, we choose the bosonic cutoff function
to be as close as possible in form to the fermionic one, \begin{equation}
R_{B}(\vec{q},k)=Z_{m}(k)\frac{(c_{B}k)^{2}-q^{2}}{4M}\,\theta(c_{B}k-q),\label{eq:sharp-boson}\end{equation}
apart from the addition of a parameter $c_{B}$, which sets the relative
scale of the fermionic and bosonic regulators, and a factor of $Z_{m}$.
The latter has the important advantage of leading to a consistent
scaling behaviour, so that all contributions to a single evolution
equation decay with the same power of $k$ for large $k$. Moreover
it also gives $a_{F}$-scaling, where all terms in a single equation
have the same dependence on $a_{F}$. Neither of these two conditions
is actually required, and we shall show below that removing such restrictions
may give us access to interesting information. 

It has been shown for bosonic systems \cite{SandM}, and it also shown
in detail in Fig.~\ref{fig:Ratio} below, that the simple two-body
truncation is insufficient. The complete set of contact interactions
in the four-body sector includes not just the dimer-dimer term $u_{2}$,
but also terms where the dimer is broken into two fermions, Eq.~
(\ref{eq:ansatz-1}). The evolution equation for any of the couplings
is described by an expansion about the energy of the bound state pole
for bosons, and half that energy for fermions, e.g., \begin{equation}
\partial_{k}\lambda=-\frac{i}{2}\,\frac{\delta^{4}\STr\left[\partial_{k}R(\Gamma^{(2)}-R)^{-1}\right]}{\delta\phi^{\dagger}\!(\mathcal{E}_{D},0)\delta\phi(\mathcal{E}_{D},0)\delta\psi^{\dagger}\!(\mathcal{E}_{D}/2,0)\delta\psi(\mathcal{E}_{D}/2,0)}.\end{equation}

\begin{figure}
\begin{centering}
\includegraphics[width=7.5cm]{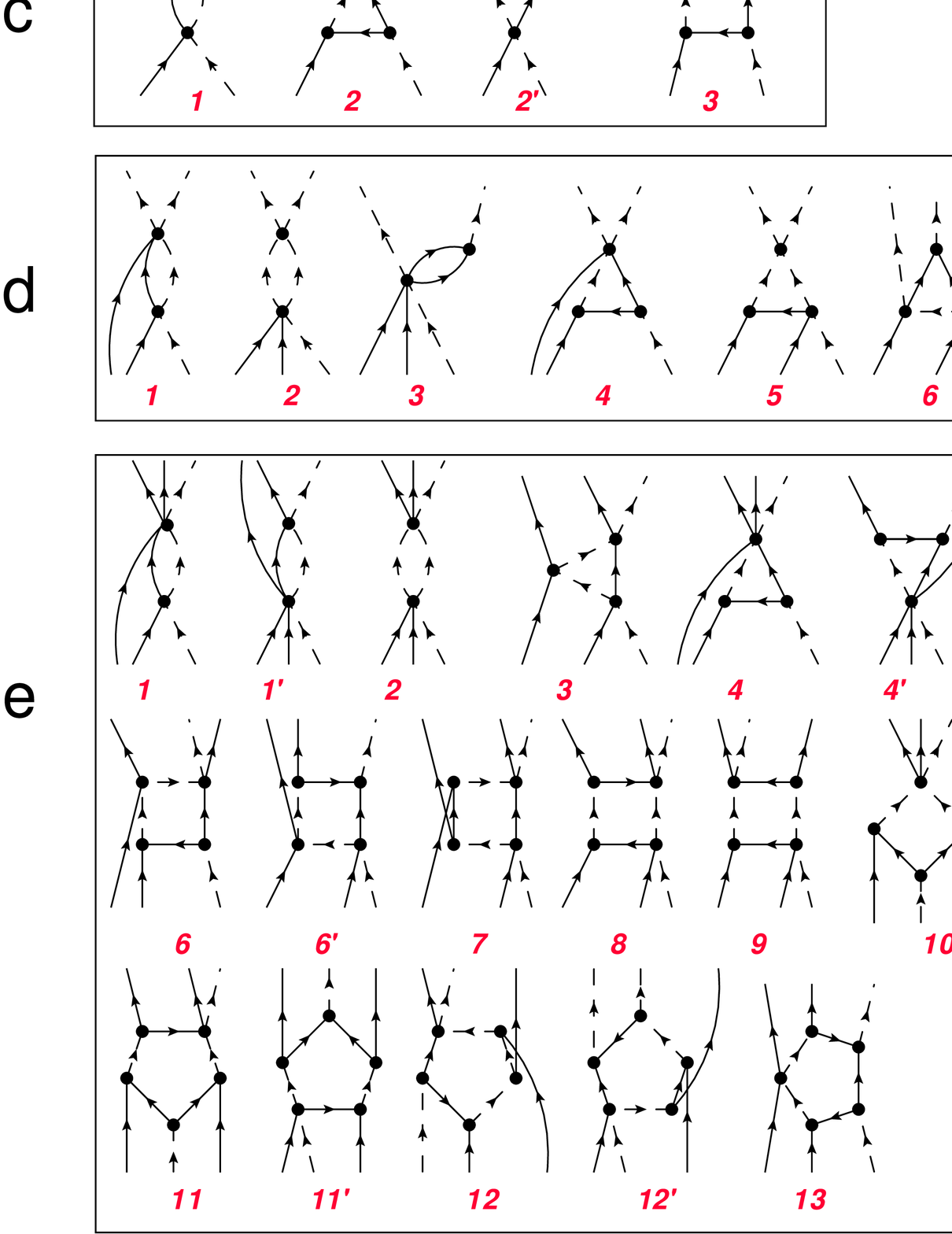} 
\par\end{centering}

\caption{{[}Colour online{]} The skeletons of the diagrams that contribute
to the evolution of the couplings. Dashed lines represent bosons,
solid lines fermions. The dots represent the interaction vertices.
In a) we show the diagrams for the evolution of $u_{1}$, $Z_{m}$
and $Z_{\phi}$ in b) for $u_{2}$, c) for $\lambda$, d) for $g'$
and e) for $\nu$.\label{fig:skeleton}}

\end{figure}

The technique to evaluate such contributions can most compactly be
written as a combination of a diagrammatic and algebraic approach.
We first evaluate the skeleton diagrams that contribute to a given
vertex. There are three distinct contributions to the running of $\lambda$,
coming from ladder, triangle and box diagrams, as shown in Fig.~\ref{fig:skeleton}.
For the other couplings, we have a large number of diagrams that can
contribute. The most difficult calculation is for $g'$, where we
have loops with up to six internal lines. Each diagram involves a
single integration over an internal four momentum. Since our cut-off
function is only momentum-dependent, i..e., is independent of the
energy variable, we can perform the energy integration by a contour
integration, enumerating the poles by solving linear algebraic equations.
The insertion of the $k$ derivative of the cut-off function on each
leg can then be achieved afterwards, by a functional derivative of
the resulting integrals with respect to the cut-off function.

The resulting equations can be written in a compact form as,\begin{widetext}\begin{align*}
\partial_{\kappa}u_{2}(\kappa) & =\frac{1}{2}I_{3,0,0}-\frac{1}{Z_{\phi}}I_{0,1,0}u_{2}(\kappa)^{2}+2I_{2,0,0}\lambda(\kappa)-2I_{1,0,0}g'(\kappa),\\
\partial_{\kappa}\lambda(\kappa) & =-\frac{1}{2}I_{2,0,1}-2I_{1,0,1}\lambda(\kappa)-2I_{0,0,1}\lambda(\kappa)^{2}\\
\partial_{\kappa}g'(\kappa) & =-2I_{1,0,1}g'(\kappa)-4I_{0,0,1}g'(\kappa)\lambda(\kappa)+\frac{-1}{Z_{\phi}}I_{0,1,0}g'(\kappa)u_{2}(\kappa)\\
 & \quad+2I_{1,1,1}u_{2}(\kappa)+2I_{2,0,1}\lambda(\kappa)+4I_{1,0,1}u_{2}(\kappa)\lambda(\kappa)+4I_{1,0,1}\lambda(\kappa)^{2}-2I_{1,0,0}\nu(\kappa)\\
-\partial_{\kappa}\nu(\kappa) & =-\left(\frac{1}{2}I_{3,1,1,}+\frac{3}{2}Z_{\phi}I_{2,1,2}\right)-4I_{1,0,1}\nu(\kappa)-8I_{0,0,1}\nu(\kappa)\lambda(\kappa)\\
 & \quad-(I_{2,1,1}+7Z_{\phi}I_{1,1,2})\lambda(\kappa)+2(I_{1,1,1}-5Z_{\phi}I_{0,1,2})\lambda(\kappa)^{2}+4I_{0,0,2}\lambda(\kappa)^{3}\\
 & \quad+2I_{1,1,1}g'(\kappa)-\frac{1}{2Z_{\phi}}I_{0,1,0}g'(\kappa)^{2}+4I_{0,1,1}g'(\kappa)\lambda(\kappa),\end{align*}
with the basic integrals

\begin{align}
I_{n_{1}n_{2}n_{3}} & =\frac{1}{2}\int_{0}^{\infty}dq'\left[\partial_{k}R_{F}(q')\frac{\delta}{\delta R_{F}(q')}+\partial_{k}R_{B}(q')\frac{\delta}{\delta R_{B}(q')}\right]_{Z_{\phi}}\nonumber \\
 & \qquad\frac{1}{(2\pi)^{3}}\int\frac{d^{3}q}{E_{\text{FR}}(q,k)^{n_{1}}E_{\text{BR}}(q,k)^{n_{2}}(E_{\text{BR}}(q,k)+Z_{\phi}E_{\text{FR}}(q,k))^{n_{3}}}.\end{align}
\end{widetext}

\section{Results}

\subsection{Full evolution}

In theory we should start integration of the evolution equations at
infinity--in practice the results are numerically independent of the
starting scale provided this is chosen to be at least $ka_{F}\simeq100$.
For $k\gg1/a_{F}$ the system is in the universal {}``scaling regime'',
and a stable fixed point, see below, governs the evolution until $k$
becomes comparable with $1/a_{F}$. 

\begin{figure}
\begin{centering}
\includegraphics[clip,width=6.5cm]{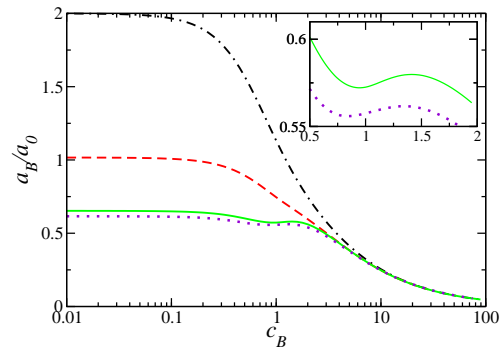} 
\par\end{centering}

\caption{{[}Colour online{]} Ratio of dimer-dimer to fermion-fermion scattering
lengths as a function of the relative scale parameter $c_{B}$. The
black (dash-dotted) curve shows results for the minimal action; the
red (dashed) curve shows the effect of adding the local three-body
term; The green (solid) curve displays the full local four-body calculation,
and is enlarged in the inset. The purple (dotted) line shows the effect
of using a smooth cut-off ($\sigma=0.5$) on the full results. \label{fig:Ratio}}

\end{figure}

In this scaling regime, we can determine the behaviour most easily
in terms of dimensionless {}``scaling variable'' $\kappa=ka_{F}$
by defining the four dimensionless functions $c_{i}(\kappa)$ \begin{align}
u_{2}(\kappa) & \to g^{4}M^{3}a_{F}^{3}\kappa^{-3}c_{0}(\kappa),\nonumber \\
\lambda(\kappa) & \to g^{2}Ma_{F}^{2}\kappa^{-2}c_{1}(\kappa),\nonumber \\
g'(\kappa) & \to g^{3}M^{2}a_{F}^{4}\kappa^{-4}c_{2}(\kappa),\nonumber \\
\nu(\kappa) & \to g^{2}Ma_{F}^{5}\kappa^{-5}c_{3}(\kappa).\end{align}
The functions $c_{i}$ satisfy a set of dimensionless differential
equations. For each level of truncation of the effective action we
can determine the resulting nontrivial fixed points, and the evolution
close to these points, as given by the anomalous dimensions $\eta_{j}$,
\[
c_{i}(\kappa)=c_{i}^{\text{fp}}+\sum_{j}a_{ij}\kappa^{\eta_{j}}.\]
For each truncation considered here we find only one stable fixed
point, see Table \ref{tab:The-coefficients-and}. There we give the
value for fixed point and anomalous dimensions for the sharp cut-off
(\ref{eq:LitimCutoff}) with the parameter $c_{B}=1$ in the bosonic
cut-off (\ref{eq:sharp-boson}). These results are somewhat dependent
on $c_{B}$, as expected, and are also not totally cut-off independent,
but seem to be reasonably stable under perturbations.

\begin{table*}
\caption{The coefficients and the eigenvalues of the stable fixed point for
various level of truncation\label{tab:The-coefficients-and}}

\centering{}\begin{tabular}{ccc}
truncation
 & fixed point & anomalous dimensions\tabularnewline
\hline
$u_{2}$ & $c_{0}=0.0656565$ & $3.1728$\tabularnewline
$u_{2},\lambda$ & $c_{0}=0.037878,c_{1}=-0.322441$ & $3.09969,3.10355$\tabularnewline
$u_{2},\lambda,g',\nu$ & $\qquad c_{0}=0.0356522,c_{1}=-0.322441,\qquad$  & $6.10943,5.15046,4.19149,3.10355$\tabularnewline
 & $c_{2}=0.102042,c_{3}=-10.488$ & \tabularnewline
\end{tabular}
\end{table*}

The dimension for $\lambda$, \begin{equation}
\eta_{\lambda}=\frac{2}{5}\sqrt{\frac{301}{3}}\approx3.10355\end{equation}
should be compared to the exact result $4.32244$, found by Griesshammer
and others\emph{ }\cite{Grie,Bir06,WeCa}. The lowest anomalous dimension
for the four-fermion sector, $4.19149$, should be compared with the
value $5.0184$ obtained numerically by Stecher and Greene \cite{StG}
(see also Ref.~\cite{ABF}). 

If we start the evolution from the stable fixed point, we can then
carefully trace this back to finite $k$. The behaviour of $a_{B}/a_{F}$
as a function of $c_{B}$ for both a sharp and a smooth cut-off with
$\sigma=0.5$, Eqs. (\ref{eq:LitimCutoff}) and (\ref{eq:SmoothCutoff}),
is presented in Fig.~\ref{fig:Ratio}. As we increase the complexity
of the truncation, we see a rapid convergence: the inclusion of $\lambda$
reduces the size of $a_{B}$ by a factor of about 2 (for small $c_{B}$),
and adding the remaining terms in the action pushes the result down
even further to agree with the Schrödinger equation results; in that
case we also see a very weak dependence on $c_{B}$ in the region
around 1. 

Note that in all cases the dominant contributions to $a_{B}/a_{F}$
for large $c_{B}$ come from the boson-loop terms in the equation
for $u_{2}$. Since these do not depend on the three-body coupling
$\lambda$, the curves approach each other. Moreover, this limit corresponds
to integrating out the fermions first, which generates a non-zero
value for $u_{2}$ at the start of the bosonic integration. In the
limit $c_{B}\rightarrow\infty$, this coupling is driven to the trivial
fixed point of the RG equations, $u_{2}=0$, since we have no terms
to cancel the linearly divergent boson-boson loop diagram and the
diagrams with three-body and four-body couplings are all too weak
to alter this behaviour.

On the other hand, the main contributions for small $c_{B}$ come
not only from the fermion loops, but also from mixed fermion-boson
loops, which appear in the equations for the many-body couplings.
In particular, the mixed boson-fermion loop diagrams containing the
fermionic cut-off contribute to the evolution of $\lambda$, $\nu$
and $g'$, even when the bosonic degrees of freedom have been integrated
out. As a result, inclusion of the three-body term $\lambda$ already
leads to a significant deviation from the mean-field result, $a_{B}/a_{F}=2$,
that persists in the limit $c_{B}\rightarrow0$. With $g'$ and $\nu$
included we see convergence close to the exact result, even when $c_{B}=0$.
We seem to have approximate convergence for a range of values of $c_{B}$,
probably best near $c_{B}=1$, but we can probably use any $c_{B}\le1.5$.
The strange results obtained for very large values of $c_{B}$ should
not be taken too seriously, since they are based on an incorrect approach:
we make the induced bosonic degrees of freedom dominate in the early
stages of evolution (large $k$). The other extreme, albeit naively
equally incorrect, actually seems to produce sensible results. This
corresponds to freezing the renormalisation of the bosonic degrees
of freedom, while still allowing an evolution of the coupling constants
driven by the evolution of the fundamental fermionic fields, suggesting
that this may be a sensible and simplifying approach. This may have
important practical consequences for calculating in the many-body
system: If we can integrate out the bosons at every stage, and only
let the fermions evolve, the calculations become much simpler. 

Arguments based on {}``optimisation'' of the cut-off function, see
Ref.~\cite{Pawlowski}, indicate that one should choose the cut-off
to try to maximise the rate of convergence for our expansion of the
action. In this case there is a stationary point for $c_{B}$ close
to 1, which agrees with the natural assumption that $c_{B}=1$, where
bosons and fermions are renormalised at the same rate is the optimal
choice.

\begin{figure}
\begin{centering}
\includegraphics[clip,width=6.5cm]{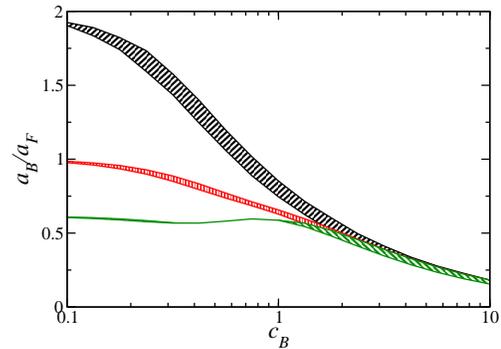} 
\par\end{centering}

\caption{{[}Colour online{]} Ratio of dimer-dimer to fermion-fermion scattering
lengths as a function of the relative scale parameter $c_{B}$. The
black (right-slanted hash) curve shows results for the minimal action;
the red (vertical hash) curve shows the effect of adding the local
three-body term; The green (left-leaning hash) curve displays the
full local four-body calculation. The width of each curve denotes
the sensitivity of the result to changes in $a_{F}$ at $a_{F}=gM/\hbar^{2}$
\label{fig:Ratio-1}}

\end{figure}

To provide a further check of convergence, it pays to use a different
form for the boson regulator, and purposefully violate both the uniform
$a_{F}$-dependence and the uniform scaling for large $k$. If we
have convergence, the results should remain independent of the cut-off,
and thus also independent of $a_{F}$ and the shape of the cut-off
function.. The simplest form we can choose uses the smooth cut-off
function (\ref{eq:SmoothCutoff}), where we renormalise the bosons
as (note the absence of $Z_{m}$)\[
R_{B}=\frac{(kc_{B})^{2}}{2m_{D}}\theta_{\sigma}(q,kc_{B}).\]
With this choice we expect the ratio $a_{B}/a_{F}$ to be a function
of $a_{F}$; we have first performed calculations at $a_{F}=1$, and
looked at the sensitivity to changes in $a_{F}$. In Fig.~\ref{fig:Ratio-1}
we see a strong dependence on $a_{F}$ for any value of $c_{B}\neq0$
for the two-body truncation, a weaker dependence for the three-body
case, and a very weak dependence for the four-body case, as long as
we consider $c_{B}<1$. For large $c_{B}\geq2$, where we let the
bosons dominate, we have universally poor result whatever the truncation,
in agreement with the discussion above.

\begin{figure}
\begin{centering}
\includegraphics[clip,width=6.5cm]{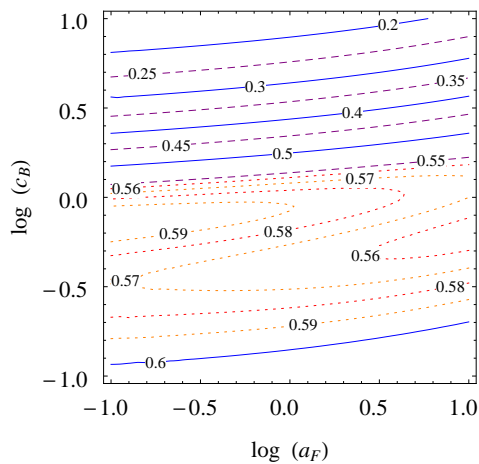} 
\par\end{centering}

\caption{{[}Colour online{]} Ratio of dimer-dimer to fermion-fermion scattering
lengths as a function of the relative scale parameter $c_{B}$, and
the logarithm of the scattering length $a_{F}$ (divided by $gM/\hbar^{2}$).
Notice the insensitivity to $a_{F}$ for $c_{B}<1$, a clear indication
of convergence. \label{fig:Ratio-2}}

\end{figure}
The convergence for the full four-body truncation can be seen even
more clearly in Fig.~\ref{fig:Ratio-2}, where we show a contour
plot of the ratio of dimer-dimer and fermion-fermion scattering lengths
as a function of $a_{F}$ and $c_{B}$. We note the excellent convergence
in a large region of the parameter space, with allows a conservative
estimate $a_{B}/a_{F}=0.58\pm0.02$, in excellent agreement with the
exact result.

\section{Discussion}

We have gathered considerable evidence for the convergence of a gradient
expansion of the quantum effective action for a system of a few dilute
fermions interacting through a pairwise attractive force. The resulting
dimers, fermion-fermion bound states, scatter in the way predicted
by exact calculations, if we expand the polarisation to second order
in momenta, and include all the local four-body terms in the fermion
and dimer fields.

Of course the expansion is only complete in terms of the number of
fields that enter the action: we have neglected non-locality of all
but the simplest terms, and have only included time derivatives to
match the momentum dependence. In principle, we can add any momentum-dependence
to any terms without problems; it is much more difficult to add energy
dependent terms; with a momentum-dependent cut-off we are limited
to first order terms only. Fortunately, it appears that we do not
need such complications! As long as we study low-energy physics, which
is exactly the situation here and in the many-body situation, that
is not too surprising. 

What does come as a surprise is that we seem to be able to fix the
bosonic fields, and have an RG flow driven by the evolution of the
fermionic fields only, while still obtaining good results. This is
probably due to the fact that the evolution of the induced dimer degrees
of freedom can be thought of as driven from the basic fermionic degrees
of freedom through the coupling constants. This requires confirmation
for finite density many-body systems. 
\begin{acknowledgments}
The work of BK is supported by the EU FP7 programme (Grant 219533).\end{acknowledgments}

\end{document}